# Accurate Analytical Modeling of Small-Size Rotary Transformers for Wound-Rotor Resolvers

*Saeed Hajmohammadi, MohammadSadegh KhajueeZadeh, Farid Tootoonchian, Senior, IEEE, and Sajjad Mohammadi, Member, IEEE*

***Abstract*— Rotary transformers are commonly used in wound-rotor resolvers to transfer excitation signals to the rotating winding without mechanical contact. In many analyses, the rotary transformer is modeled as an ideal transformer, where the voltage transfer ratio is assumed to be equal to the turns ratio. However, in miniature rotary transformers used in compact resolver systems, leakage inductance can become comparable to the magnetizing inductance due to reduced core dimensions and unavoidable air gaps, leading to deviations from the ideal voltage transfer behavior. This paper presents an accurate equivalent circuit model for miniature rotary transformers employed in wound-rotor resolvers. The proposed model analytically derives the magnetizing and leakage inductances using a magnetic equivalent circuit that accounts for flux fringing and air-gap effects. The model is validated through three-dimensional finite element analysis and experimental measurements on a fabricated prototype under both no-load and resolver excitation conditions. The results demonstrate improved prediction accuracy of the secondary voltage compared with conventional models, enabling more reliable characterization of excitation transfer in compact resolver systems.**

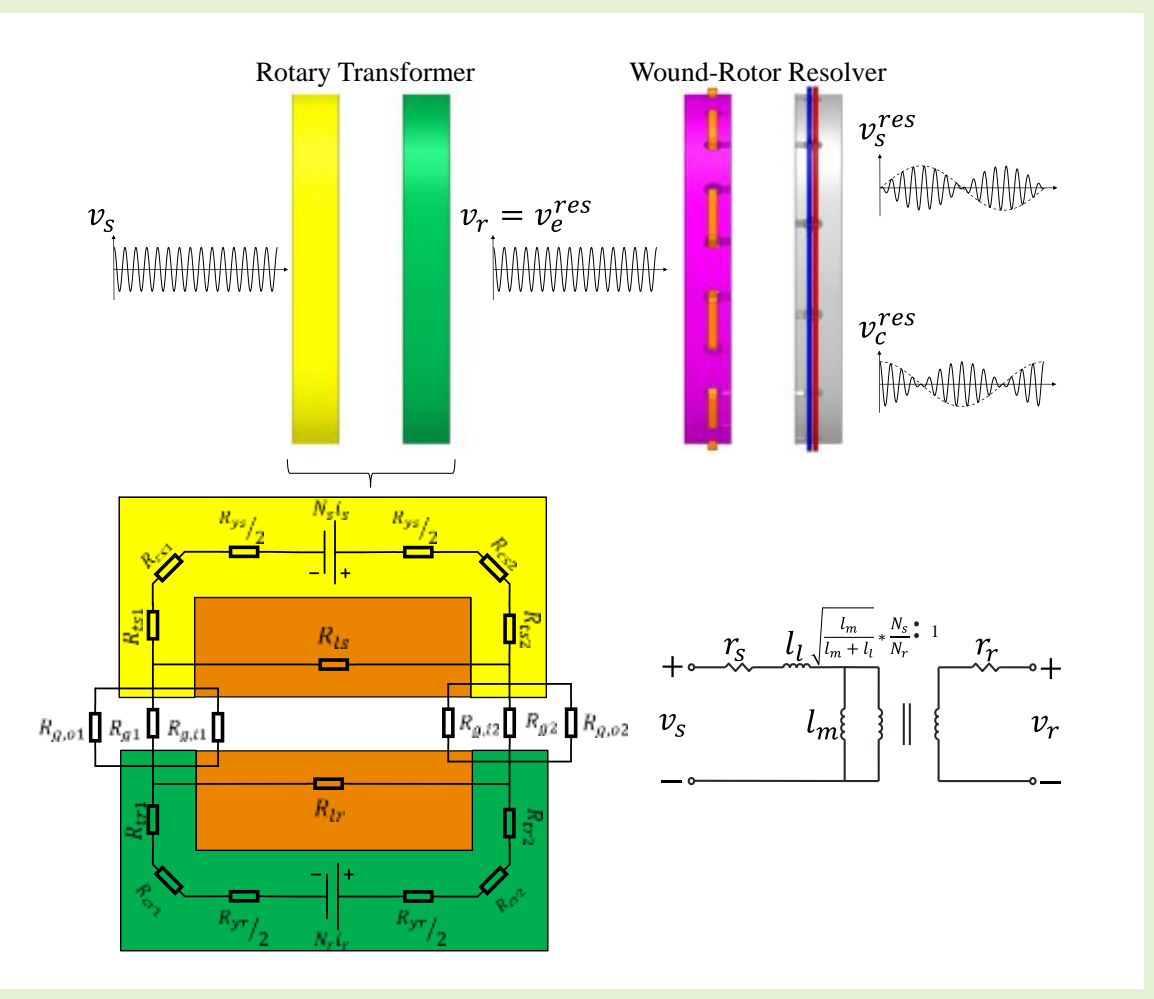

## I. Introduction

Accurate position sensing is essential in modern motion control systems such as industrial servo drives, robotics, electric vehicles, aerospace actuators, and precision manufacturing equipment. The performance and reliability of these systems largely depend on the accuracy and robustness of the employed position sensors, particularly in harsh environments involving high temperatures, mechanical vibration, and electromagnetic interference. Among available sensing technologies, resolvers and synchros are widely used in industrial and safety-critical applications due to their rugged structure, long operational lifetime, and strong tolerance to harsh environmental conditions [1]–[2]. Compared with synchros, resolvers feature simpler signal processing and lower winding complexity, making them widely preferred in many contemporary motion control systems. A resolver typically includes one excitation and two signal windings. When excited by a high-frequency sinusoidal reference signal, the resolver generates two quadrature amplitude-modulated output signals [3]. Using these signals, together with the reference excitation, the rotor position is subsequently reconstructed by a resolver-to-digital converter (RDC). Therefore, the fidelity of the excitation transfer and the quality of the resulting modulated signals have a direct impact on the achievable position-estimation accuracy.

Resolvers are commonly categorized into variable-reluctance and wound-rotor types [4]–[5]. While variable-reluctance structures offer mechanical simplicity, wound-rotor resolvers generally provide higher signal quality and accuracy, especially in low pole-count designs, and are therefore widely used in high-precision and long-life applications [6]. In wound-rotor resolvers, the excitation winding is placed on the rotor, requiring a mechanism to transfer excitation power from the stator. Although slip rings were used in early implementations, their mechanical contact and wear limit reliability [7]. To achieve contactless and maintenance-free operation, rotary transformers are commonly

Saeed Hajmohammadi, and Farid Tootoonchian are with the Electrical Engineering Department, Iran University of Science and Technology, Tehran 11155-8639, Iran (e-mail: s.hajmohammadi1994@gmail.com; tootoonchian@iust.ac.ir). MohammadSadegh KhajueeZadeh is with the Department of Electrical and Computer Engineering, University of British Colombia, Vancouver V6T 1Z4, BC, Canada (e-mail: mohammadkhajuee@ece.ubc.ca).

Sajjad Mohammadi is with the Department of Electrical and Computer Engineering, University of Alberta, Edmonton T6G 2H5, AB, Canada (e-mail: sajjad.mohammadi@ualberta.ca).

employed to inductively transfer excitation voltage while preserving magnetic coupling during rotation, significantly enhancing system robustness and durability [8].

Extensive research has been devoted to improving the electromagnetic design and structural configuration of wound-rotor resolvers. Various winding topologies have been investigated to enhance signal quality and reduce position error. In [9], different winding arrangements were analyzed, and distributed winding structures were identified as promising candidates for performance improvement. Zare in [10] compared toroidal and tooth-wound excitation configurations and demonstrated that toroidal windings can significantly reduce sensing error. Similarly, Naderi et al. in [11] implemented optimized winding configurations in both stator and rotor structures to develop a high-accuracy wound-rotor resolver. Other studies have explored alternative structural concepts, including printed-circuit-based integrated designs [12], limited-angle high-precision resolvers [13], hybrid structures capable of simultaneous rotational and linear sensing [14], high pole-count resolvers with non-overlapping windings [15], and multi-pole signal generation within conventional resolver frameworks [16].

Despite these advancements, comparatively limited attention has been paid to the electrical modeling of the rotary transformer itself, even though it directly governs excitation transfer in wound-rotor configurations. Several studies have investigated structural modifications of rotary transformers to reduce leakage flux or phase displacement error [17]–[18]. However, most existing analyses treat the rotary transformer as an ideal transformer, assuming that the voltage transfer ratio is equal to the turns ratio. While this assumption is generally acceptable for conventional transformers with negligible leakage inductance, it may lead to inaccuracies in miniature rotary transformers used in compact resolver systems. In small-scale structures, reduced core dimensions and unavoidable air gaps cause the leakage inductance to become comparable to the magnetizing inductance. Under such conditions, the classical turns-ratio-based voltage transfer assumption is no longer valid. The resulting deviation in excitation voltage transfer can introduce amplitude errors, gain imbalance between the sin/cos outputs, demodulation inaccuracies, and increased uncertainty in position estimation. Therefore, an accurate equivalent circuit model that captures both magnetizing and leakage effects is essential for reliable prediction of excitation behavior and improved sensing performance. To address this issue, this paper presents an accurate equivalent circuit model for miniature rotary transformers employed in wound-rotor resolvers. The main contributions of this work are summarized as follows:

- Analytical investigation of the non-ideal voltage transfer ratio in miniature rotary transformers, demonstrating its deviation from the turns ratio due to significant leakage inductance.
- Derivation of magnetizing and leakage inductances using a magnetic equivalent circuit that incorporates flux fringing and magnetic potential drops across the air gap.
- Validation of the proposed analytical formulations using three-dimensional finite element analysis (3D FEA) under different air-gap conditions.
- Experimental verification using a fabricated rotary transformer prototype under both no-load and resolver excitation conditions, demonstrating improved prediction accuracy compared with conventional models.

The remainder of this paper is organized as follows. Section II reviews conventional transformer equivalent circuit model and illustrates the voltage transfer ratio mismatch using frequency-response analysis. Section III introduces the proposed magnetic modeling approach for accurate inductance estimation and validates the results using 3D FEA. Section IV presents the experimental setup and measurement results under both no-load and resolver excitation conditions. Finally, Section V concludes the paper.

## II. Proposed Sensor System Description and Conventional Modeling

In The overall configuration of a typical wound-rotor resolver integrated with a rotary transformer is illustrated in Fig. 1. The rotary transformer serves as the contactless interface for excitation transfer. When the stationary primary winding of the rotary transformer is energized by a high-frequency carrier signal, a voltage is induced in the rotating secondary winding through electromagnetic induction. This induced voltage then acts as the excitation source for the resolver's rotor winding. Consequently, the resolver signal windings generate two amplitude-modulated outputs (sine and cosine), from which the angular position is extracted by using these signals and the reference excitation.

### A. Conventional Equivalent Circuit Model

Since the accuracy of position estimation is directly influenced by the amplitude and phase characteristics of the resolver signals, an accurate electrical model of the rotary transformer is essential [18]. Fig. 2 shows the conventional equivalent circuit model widely used for rotary transformers in previous studies. In this model, $r_s$ and $r_r$ represent the resistances of the stator and rotor windings, respectively. The parameters $L_l$ and $L_m$ denote the total leakage inductance and the magnetizing inductance, while $n$ represents the voltage transfer ratio. Typically, these parameters are derived from the stator self-inductance ($L_{ss}$), rotor self-inductance ($L_{rr}$), and the mutual inductance ($L_{sr}$) between the two windings as follows:

$$k = \sqrt{l_{s,r} l_{r,s} / l_{ss} l_{rr}} = \sqrt{l_m / (l_m + l_l)} \quad (1)$$

$$n = l_{s,r} / l_{rr} = k / \sqrt{l_{rr} / l_{ss}} \quad (2)$$

$$l_m = n l_{s,r} \quad (3)$$

$$l_l = l_{ss} - l_m \quad (4)$$

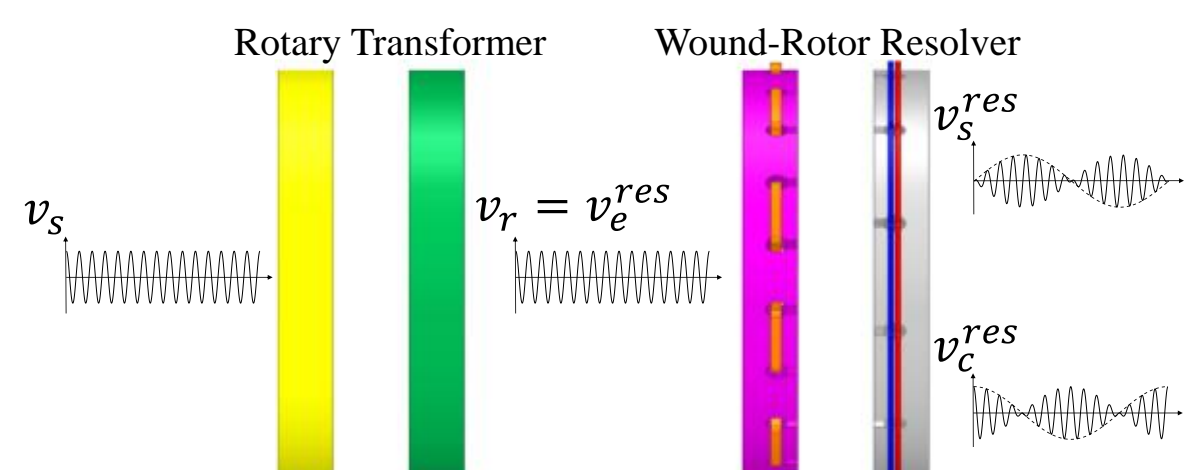

Fig. 1. The schematic of a conventional wound-rotor resolver with rotary transformer

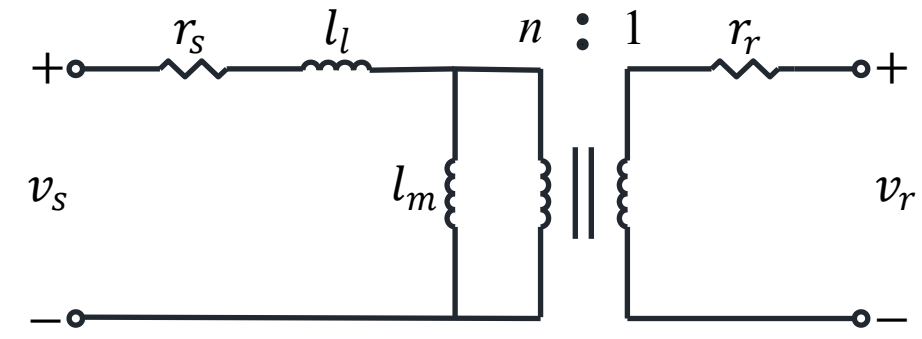

Fig. 2. Conventional equivalent circuit model of rotary transformer.

## B. Analysis of Voltage Transfer Ratio Discrepancy

In existing literature, the voltage transfer ratio of the rotary transformer is commonly assumed to be equal to its physical turns ratio ($N_s/N_p$). However, this assumption fails for miniature rotary transformers utilized in compact resolver systems. To investigate this discrepancy, two rotary transformers with different scales, a large-size design and a miniature design, are analyzed. Fig. 3 and Table I show the geometry and geometrical dimension data of this configuration for large- and small-size transformers.

Using finite element method (FEM) simulations, the self and mutual inductances were extracted, and the corresponding equivalent circuit parameters were calculated, as presented in Table II. According to the data in Table II, it is evident that the voltage transfer ratio deviates significantly from the turns ratio in the miniature transformer, whereas they are nearly identical in the large-scale version. This discrepancy arises because, in small-scale structures, the leakage inductance ($L_l$) becomes comparable in magnitude to the magnetizing inductance ($L_m$) due to the reduced core volume and relatively large air gaps.

For the equivalent circuit shown in Fig. 2, the rotor-to-stator voltage ratio can be expressed as (5).

$$H_{Tr} = \frac{v_r}{v_s} = \frac{l_m z_l' s / n}{l_l l_m s^2 + \left(r_s l_m + (l_m + l_l)(z_l' + r_r')\right)s + r_s (z_l' + r_r')} \quad (5)$$

Based on this formulation, the impact of the aforementioned modeling error is further illustrated through the frequency-response (Bode) analysis shown in Fig. 4. In this analysis, the transformer load $Z_L$ is modeled as a 10 Ω resistive load, and $s = j\omega$. As observed in the Bode plots in Fig. 4(a), using the ideal turns ratio leads to a substantial miscalculation of the output voltage amplitude for the miniature transformer. In contrast, for the large-scale transformer (Fig. 4(b)), the leakage effect is negligible, and the ideal assumption remains valid. This confirms that for high-precision sensing in compact applications, a more sophisticated modeling approach that accounts for these non-ideal effects is required.

# III. Proposed Method

In this section, an analytical approach based on the magnetic equivalent circuit (MEC) is developed to model the rotary transformer. The formulation of the MEC model and the derivation of the relevant electrical parameters are presented in the following subsections.

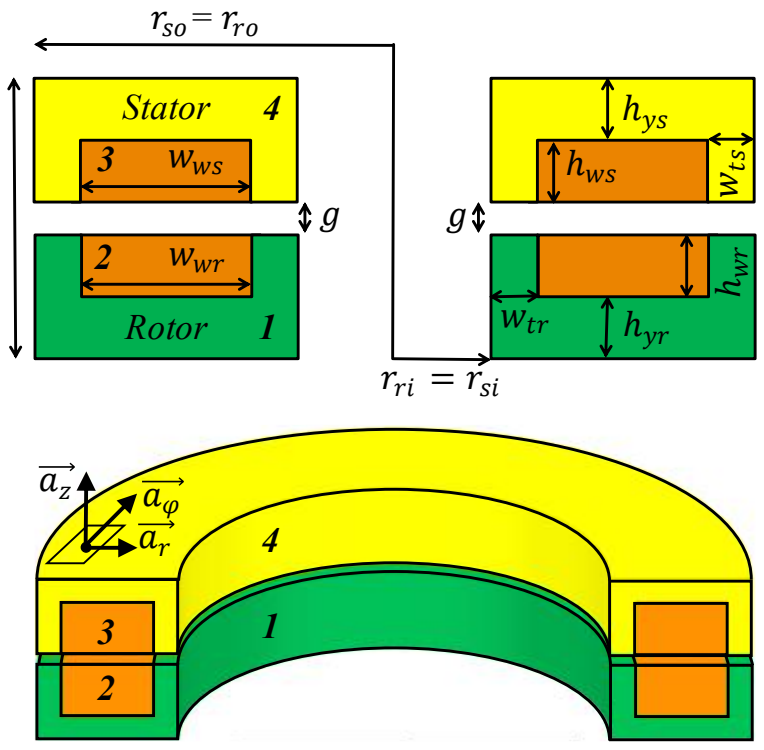

Fig. 3. Details of transformer geometry: (1) rotor core, (2) rotor winding, (3) stator winding, and (4) stator core.

TABLE I
THE GEOMETRICAL DIMENSION DATA

| Parameters | Axial Small-Size [19] | Axial Large-Size [20] |
|---|---|---|
| $h_{ws}/h_{wr}$ | 3.5mm/3.5mm | 7.8mm/7.8mm |
| $w_{ws}/w_{wr}$ | 5.5mm/5.5mm | 35.6mm/35.6mm |
| $r_{si}/r_{ri}$ | 13mm/13mm | 17mm/17mm |
| $r_{so}/r_{ro}$ | 21.5mm/21.5mm | 67.6mm/67.6mm |
| g | 0.6mm | 0.6mm |
| $w_{ts}/w_{tr}$ | 1.5mm/1.5mm | 7.5mm/7.5mm |
| $h_{ys}/h_{yr}$ | 1.5mm/1.5mm | 7.5mm/7.5mm |
| turn number ratio | 1 | 1 |

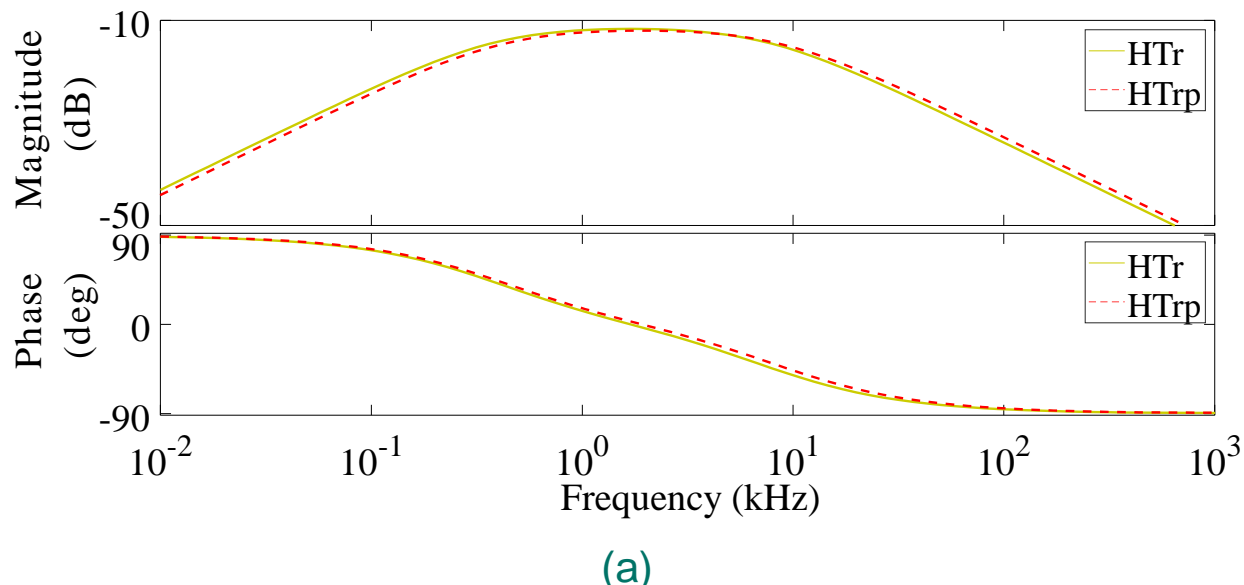

(a)

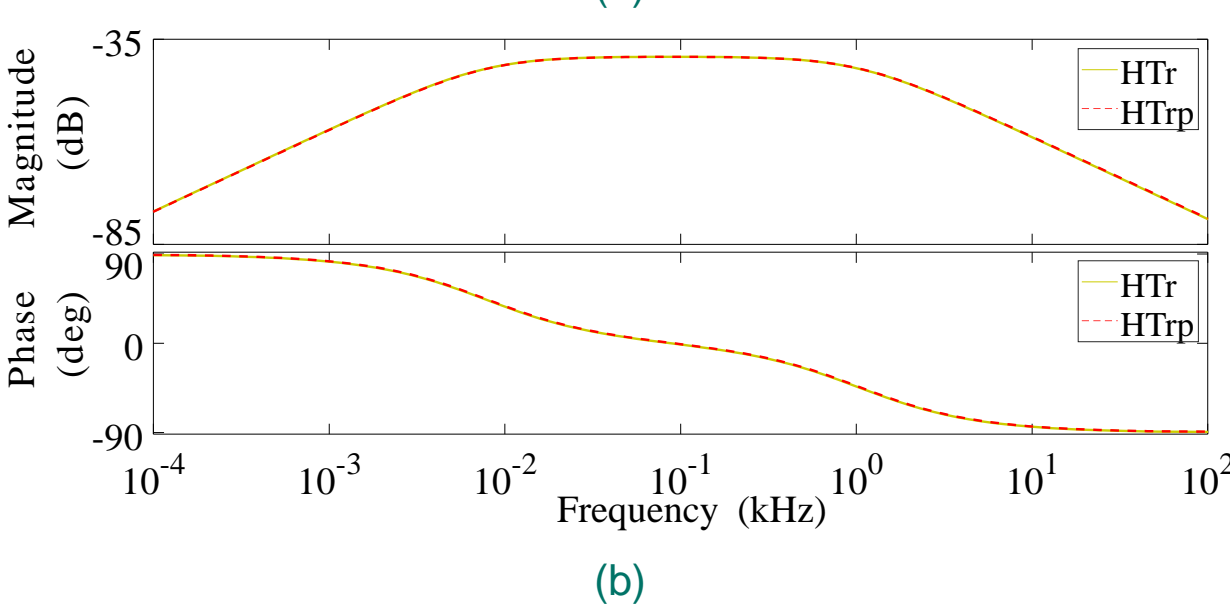

(b)

Fig. 4. Bode diagrams for transformer circuit model in Fig. 1a, where $H_{Tr}$ denotes exact transformer ratio, and $H_{Trp}$ denotes turn number ratio: (a) small-size transformer, and (b) large-size transformer.

TABLE II
DATA PARAMETERS OF CIRCUIT MODELS

| Model Parameters | Axial Small-Size | Axial Large-Size |
|---|---|---|
| $l_{ss}$ | 3.322mH | 4.693H |
| $l_{sr}$ | 2.968mH | 4.625H |
| $l_{rr}$ | 3.348mH | 4.695 H |
| $l_l$ | 0.6915mH | 0.137H |
| $l_m$ | 2.631mH | 4.556H |
| $n$ | 0.887 | 0.985 |
| *turn number ratio* | 1 | 1 |

### A. Magnetic Equivalent Circuit

Magnetic models define the correlation between the geometry of a transformer and its circuit model; a concentration on magnetization and leakage inductances.

#### 1) *Magnetization Inductance*

Fig. 5 is the magnetic circuit, which takes into account the fringe reluctance. In this regard, the magnetization inductance ($l_m$) can be written as follows, where $R_m$ is the reluctance of the main magnetic flux, and $R_{ts}$, $R_{tr}$, $R_{ys}$, $R_{yr}$, $R_{cs}$, and $R_{cr}$ are the tooth, yoke, and corner reluctances of the stator and rotor. Here, the indices of $t$, $y$, and $c$ symbolize the tooth, yoke, and corner. Moreover, $R_{g,eq}$ denotes the reluctance of airgap flux, and $N_s$ is the number of turns on the stator side.

According to (6), in addition to $R_g$, which accounts for the airgap flux in a straight line, $R_{g,o}$ and $R_{g,i}$ are also regarded to account for the fringe effect. In general, (7) is the core of the next calculations, where $dl$, $H$, $B$, and $\mu$ are in order an increment, magnetic field strength, magnetic flux density, and the medium's permeability. In Fig. 6, flux tubes are also shown, which form the basis of integrals in (7)-(17).

$$R = \frac{mmf}{flux} = \frac{\int \vec{H} \cdot \vec{dl}}{\iint \vec{B} \cdot \vec{dS}} = \int \frac{dl}{\mu S} \quad (7)$$

According to Cylindrical coordinate system $(r, \varphi, z)$ in Fig. 3 and Fig. 6, $R_{ys}$ will be written as an instance below, where the increment is in the direction of $\vec{a_r}$.

$$R_{ys} = \int_{r_{si}+w_{ts}}^{r_{so}-w_{ts}} \frac{dr}{\mu 2\pi r h_{ys}} = \ln\left(\frac{r_{so} - w_{ts}}{r_{si} + w_{ts}}\right) / \mu 2\pi h_{ys} \quad (8)$$

In the same manner, other reluctances can be written as:

$$R_{yr} = \ln\left(\frac{r_{ro} - w_{tr}}{r_{ri} + w_{tr}}\right) / \mu 2\pi h_{yr} \quad (9)$$

$$R_{ts1} = h_{ws} / \mu\pi\left((r_{si} + w_{ts})^2 - {r_{si}}^2\right) \quad (10\text{-}1)$$

$$R_{ts2} = h_{ws} / \mu\pi\left({r_{so}}^2 - (r_{so} - w_{ts})^2\right) \quad (10\text{-}2)$$

$$R_{tr1} = h_{wr} / \mu\pi\left((r_{ri} + w_{tr})^2 - {r_{ri}}^2\right) \quad (11\text{-}1)$$

$$R_{tr2} = h_{wr} / \mu\pi\left({r_{ro}}^2 - (r_{ro} - w_{tr})^2\right) \quad (11\text{-}2)$$

$$R_{g1} = g / \mu_0\pi\left((r_{ri} + w_{ts})^2 - {r_{ri}}^2\right) \quad (12\text{-}1)$$

$$R_{g2} = g / \mu_0\pi\left({r_{ro}}^2 - (r_{ro} - w_{tr})^2\right) \quad (12\text{-}2)$$

$$R_{cs,r} = \frac{1}{4\mu\left((r_{so} - w_{ts}) \ln\left(\frac{r_{so}}{r_{so} - w_{ts}}\right) + w_{ts}\right)} \quad (13)$$

To calculate the fringe effect in a magnetic circuit, (7) remains effective, but with a change in $S$ and $dl$ for $R_{sh}$.

$$R_{sh,o} = \left(\int_{\frac{g}{2}}^{r_{f,o}-\frac{g}{2}} \frac{\mu_0 L_x}{\pi r} dr\right)^{-1} \quad (14\text{-}1)$$

$$R_{sh,i} = \left(\int_{\frac{g}{2}}^{r_{f,i}-\frac{g}{2}} \frac{\mu_0 L_x}{\pi r} dr\right)^{-1} \quad (14\text{-}2)$$

Where $L_x$ is different between the inner and outer radius of the transformer, as shown in Fig. 7. Furthermore, in accordance with [21], $R_f$ is written as below:

$$R_{f,io} = 1/0.26\mu_0 L_x \quad (15)$$

According to (15), $r_f$ changes with geometry and location of the fringes. Thus, in [22] an effort was made to address this challenge through the radius definition of fringe fluxes on both the outer and inner walls of the transformer, regarding 3D-FEA as a metric of exactness. In agreement with [22], the outer radius ($r_{f,o}$) should be chosen to be the same as (16-1). For the inner fringe, which is much smaller than the outer one, it is better to choose the inner radius ($r_{f,i}$) as written in (16-2).

$$r_{f,o} = h - (h_{wr} + h_{yr}) \quad (16\text{-}1)$$

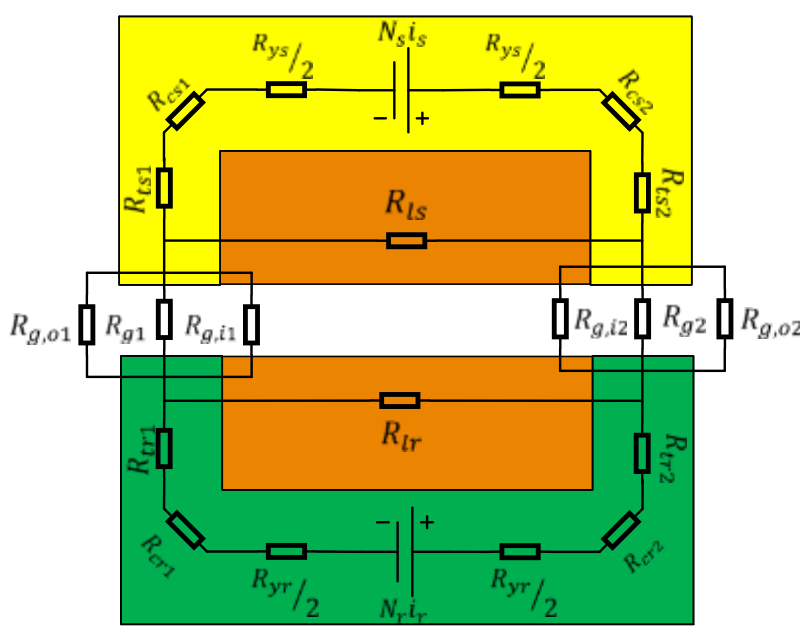

Fig. 5. Magnetic circuit of the rotary transformer.

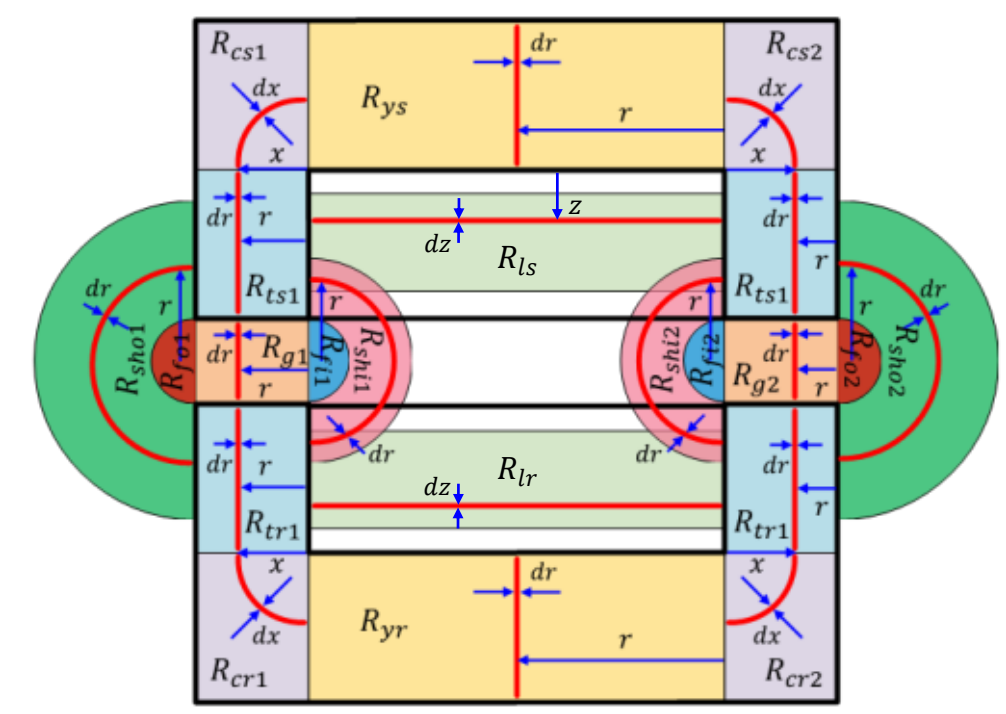

Fig. 6. Paths of integrations and flux tubes of the rotary transformer.

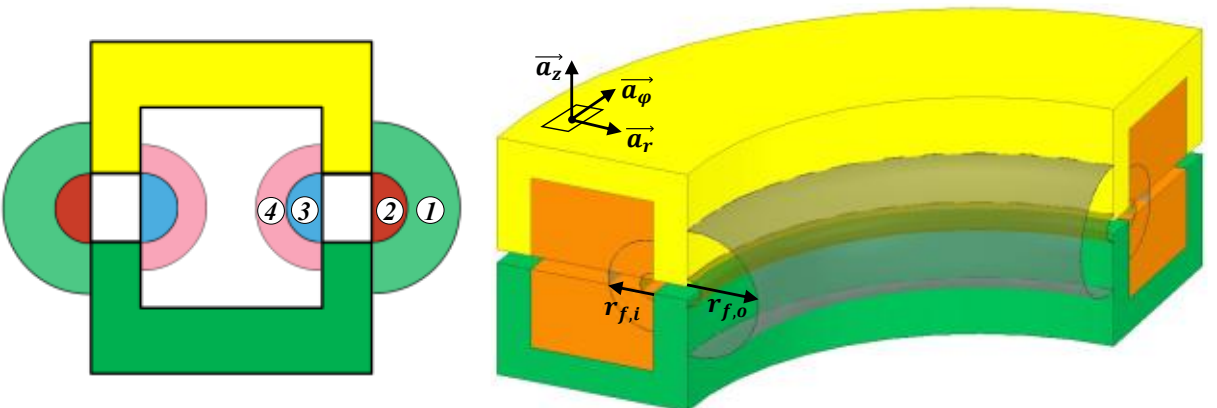

Fig. 7. Modeling the airgap fringe reluctances includes (1) $R_{sh,o}$, (2) $R_{f,o}$, (3) $R_{sh,i}$, and (4) $R_{f,i}$. It is worth to mention that $L_x$ differs between outer and inner regions, where it is $2\pi r_i$ and $2\pi(r_i + w_{ts})$, respectively. Analogous treatment can be regarded to other regions.

$$l_m = \frac{N_s^2}{R_m}$$
$$R_m = R_{ys} + R_{yr} + R_{ts1} + R_{tr1} + R_{cs1} + R_{cr1} + R_{g,eq1} + R_{ts2} + R_{tr2} + R_{g,eq2} + R_{cs2} + R_{cr2}$$
$$R_{g,eq} = R_{g,i}R_g R_{g,o} / R_{g,i}R_g + R_{g,o}\left(R_{g,i} + R_g\right)$$
$$R_{g,x} = R_{f,x}R_{sh,x} / R_{f,x} + R_{sh,x} \quad (6)$$

$$r_{f,i} = \min\left(\left(h - \left(h_{wr} + h_{yr}\right) - h_{ys}\right), \frac{w_{ws}}{2}\right) \quad (16\text{-}2)$$

According to Figs. 6 and 7, $R_{g,o}$ and $R_{g,i}$ are as shown below:

$$R_{g,o2} = 1/2\mu_0 r_{si}\left(0.26\pi + \ln\frac{r_{f,o}-\frac{g}{2}}{\frac{g}{2}}\right) \quad (17\text{-}1)$$

$$R_{g,i2} = 1\Big/2\mu_0\left(r_{si} + w_{ts}\right)\left(0.26\pi + \ln\frac{r_{f,i}-\frac{g}{2}}{\frac{g}{2}}\right) \quad (17\text{-}2)$$

$$R_{g,o1} = 1\Big/2\mu_0 r_{so}\left(0.26\pi + \ln\frac{r_{f,o}-\frac{g}{2}}{\frac{g}{2}}\right) \quad (17\text{-}3)$$

$$R_{g,i1} = 1\Big/2\mu_0\left(r_{so} - w_{ts}\right)\left(0.26\pi + \ln\frac{r_{f,i}-\frac{g}{2}}{\frac{g}{2}}\right) \quad (17\text{-}4)$$

### 2) *Leakage Inductance*

When magnetic flux links one winding in a transformer, but not the other, such flux contributes to leakage inductance. According to Fig. 8(a), in regions I and III, the flux leakage lines almost seem as aligned lines within the windows, with a height that is the same as the window width of the transformer ($w_w$). In region II, due to the air gap effect, the flux leakage lines traverse a longer length than the window width of the transformer, which is in contrast to what was taken in [23], as shown in (22) and Fig. 8(a). If we treat it in the same manner as what was shown in [23], the magnetic model will not have enough exactness in the face of leakage inductance, especially in larger air gaps. In other words, the leakage inductance will rise more than what is happening in 3D-FEM. Owing to the lack of an exact estimation of $S$ for flux leakages, (7) will not work here. Accordingly, we can define leakage inductance as an integral of magnetic energy density over the volume of the winding in the transformer window ($v_w$), as shown below. In this case, the leakage inductance, as in Fig. 1, is an integrated term on one side, either the stator or rotor. Here, $I_s$ denotes the stator current.

$$\frac{1}{2} l_l (I_s)^2 = \iiint_0^{v_w} \frac{1}{2} BH\, dv = \iiint_0^{v_w} \frac{1}{2}\mu_0 H^2 dv \quad (18)$$

If we define $H$ as follows, leakage inductance is written as shown in (20). Here, $\Psi(z)$ denotes the Magnetomotive Force (MMF).

$$N_s I_s = N_r I_r \quad (19\text{-}1)$$

$$\oint \vec{H}\,.\,\vec{dl} = \Psi(z) \quad (19\text{-}2)$$

$$l_l = \frac{\mu_0}{(I_s)^2}\iiint_0^{v_w}\left(\frac{\Psi(z)}{l(z)}\right)^2 dv \quad (20)$$

According to $N_s I_s = N_r I_r$, the MMF distribution across the cross-section of the stator- and rotor-side windings is as shown in Fig. 8(b).

$$\Psi_{(z)-Region\ I} = \frac{N_s I_s}{h_{wr}}\left(z - h_{yr}\right) \quad (21\text{-}1)$$

$$\Psi_{(z)-\ Region\ II} = N_s I_s \quad (21\text{-}2)$$

$$\Psi_{(z)-\ Region\ III} = \frac{N_s I_s}{h_{ws}}\left(-z + h_{yr} + h_{wr} + g + h_{ws}\right) \quad (21\text{-}3)$$

Where, region I is between $h_{yr}$ and $h_{yr} + h_{wr}$, region II is between $h_{yr} + h_{wr}$ and $h_{yr} + h_{wr} + g$, and region III is between $h_{yr} + h_{wr} + g$ and $h_{yr} + h_{wr} + g + h_{ws}$. According to Fig. 8(b), $l(z)$ is written as below:

$$l_{(z)-Region\ I} = w_{ws} \quad (22\text{-}1)$$

$$l_{(z)-\ Region\ II} = w_{ws} + \frac{\pi g}{2} \quad (22\text{-}2)$$

$$l_{(z)-\ Region\ III} = w_{ws} \quad (22\text{-}3)$$

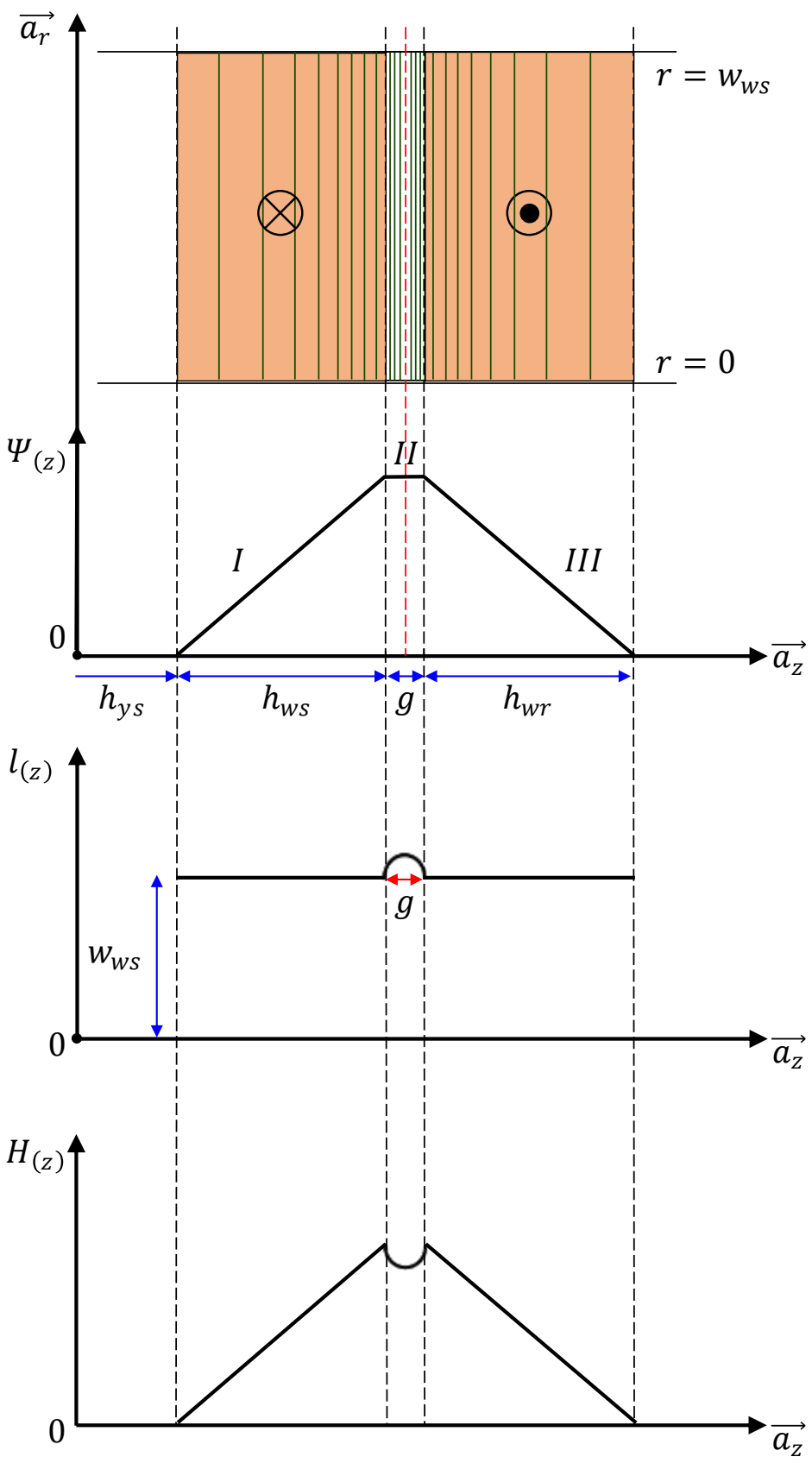

Fig. 8. (a) Flux leakage lines, and (b) leakage magnetomotive force (MMF) diagram and magnetic field strength.

If we substitute (21) and (22) into (20), we obtain leakage inductance as written (23).

$$l_l = \mu_0 \pi {N_s}^2\left(r_{si} + r_{so}\right)\left(\frac{h_{wr}}{3w_{ws}} + \frac{h_{ws}}{3w_{ws}} + \frac{g}{w_{ws} + \frac{\pi g}{2}}\right) \quad (23)$$

## B. *Accuracy of MEC*

According to the nature of the airgap, a larger airgap results in lower magnetization inductance and larger leakage inductance. Therefore, we should evaluate the sensitivity of the magnetic model to different airgap sizes. In this context, an airgap range of 0.4 mm to 4 mm was chosen, and 3D-FEA was regarded as the reference. In Fig. 9, the magnitudes of magnetic flux density and magnetic field strength along with the magnetic flux lines are shown. Moreover, the schematic of the mesh grid and its quality is shown in Fig. 10. The results shown in Figs. 11(a) and (b) indicate a high level of agreement between the suggested magnetic model and 3D-FEA for both inductances (Magnetization and Leakage). Thus, we can claim that the suggested magnetic model has more than 95% accuracy with an almost constant error in different airgaps.

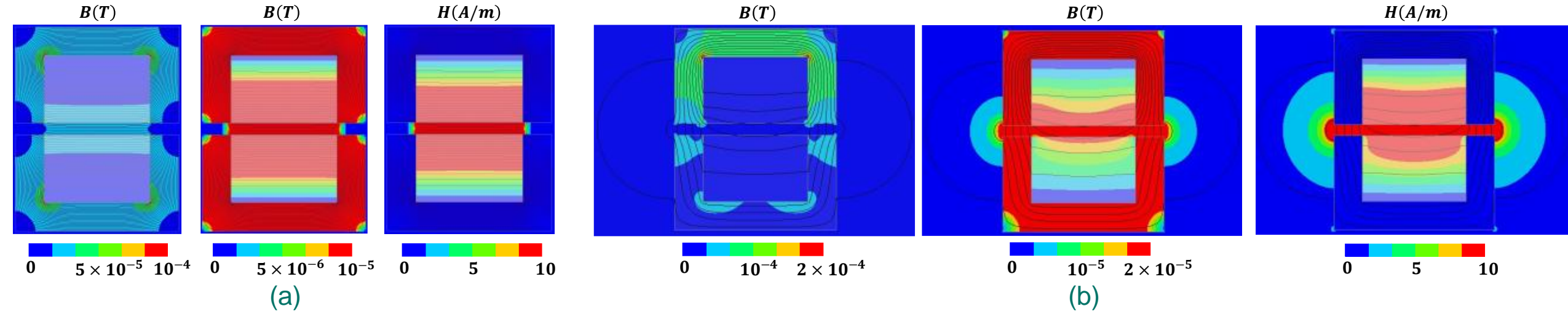

Fig. 9. The magnetic flux density and magnetic field strength for (a) $I_s = I_r$ and (b) $I_s > I_r$.

### C. Transformer Ratio Adjustment

According to the circuit shown in Fig. 1, the transformer ratio has been written as below:

$$n = \sqrt{l_{s,r}l_{r,s}/l_{ss}l_{rr}} \ / \sqrt{l_{rr}/l_{ss}} \qquad (24)$$

If we assume that $1/\sqrt{l_{rr}/l_{ss}}$ is the same as the turn number ratio, $(N_s/\ N_r)$, as shown in Fig. 1b, and $\sqrt{\frac{l_{s,r}l_{r,s}}{l_{ss}l_{rr}}} = \sqrt{\frac{l_m}{(l_m+l_l)}}$ as written in (1), then we can derive (25) as below:

$$n \cong \sqrt{\frac{l_m}{(l_m+l_l)}} \times (N_s/\ N_r) \qquad (25)$$

In Table III, exact and adjusted transformer ratios and turn number ratios were written for different configurations according to (25). It is evident that the suggested adjustment in (25) has satisfactory accuracy. Moreover, the Bode diagrams shown in Fig. 12 also confirm this conclusion.

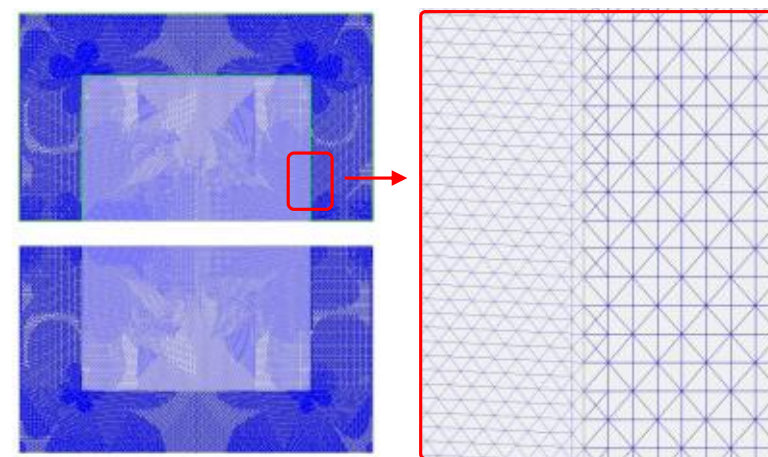

Fig. 10. The schematic of the mesh grid.

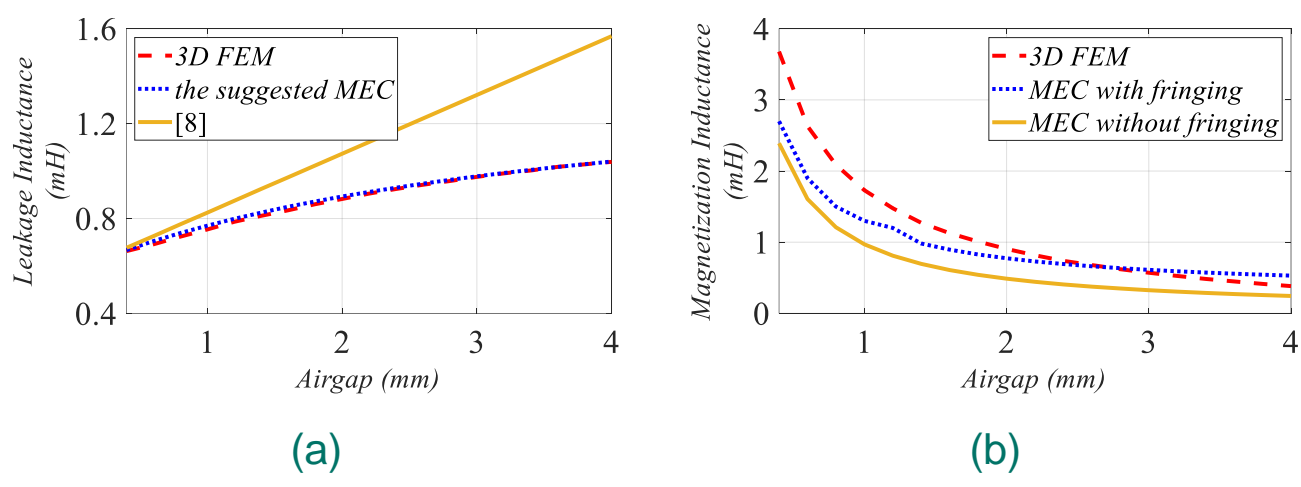

Fig. 11. Leakage inductances from 3D-FEA, the investigated methodology in [8], and the suggested MEC in this study. (b) Magnetization inductances from 3D-FEA, MEC with fringing effect, and MEC without fringing effect.

TABLE III
COMPARISON OF EXACT AND ADJUSTED TRANSFORMER RATIOS FOR DIFFERENT TURN NUMBER RATIOS

| Model Parameters | Turn number ratio | | |
|---|---|---|---|
| | 0.5 | 1 | 2 |
| $l_{ss}$ | 830.567$\mu$H | 3.322mH | 3.322mH |
| $l_{sr}$ | 1.484mH | 2.968mH | 1.484mH |
| $l_{rr}$ | 3.348mH | 3.348mH | 836.943$\mu$H |
| $n$(exact) | 0.443 | 0.887 | 1.773 |
| $n$(adjusted) | 0.445 | 0.9 | 1.780 |

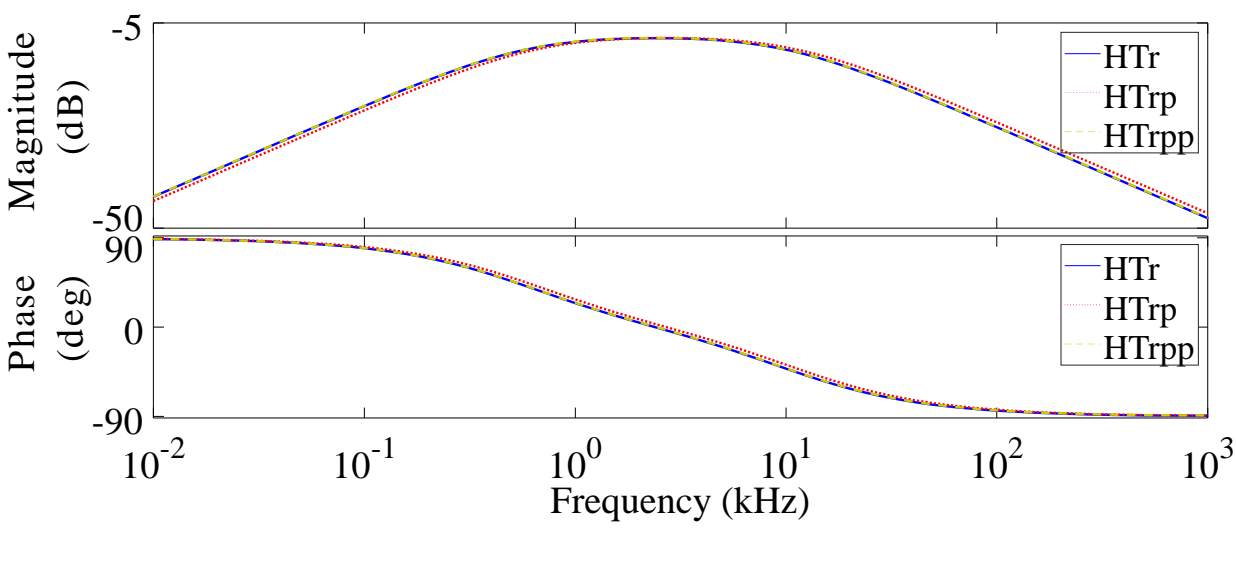

(a)

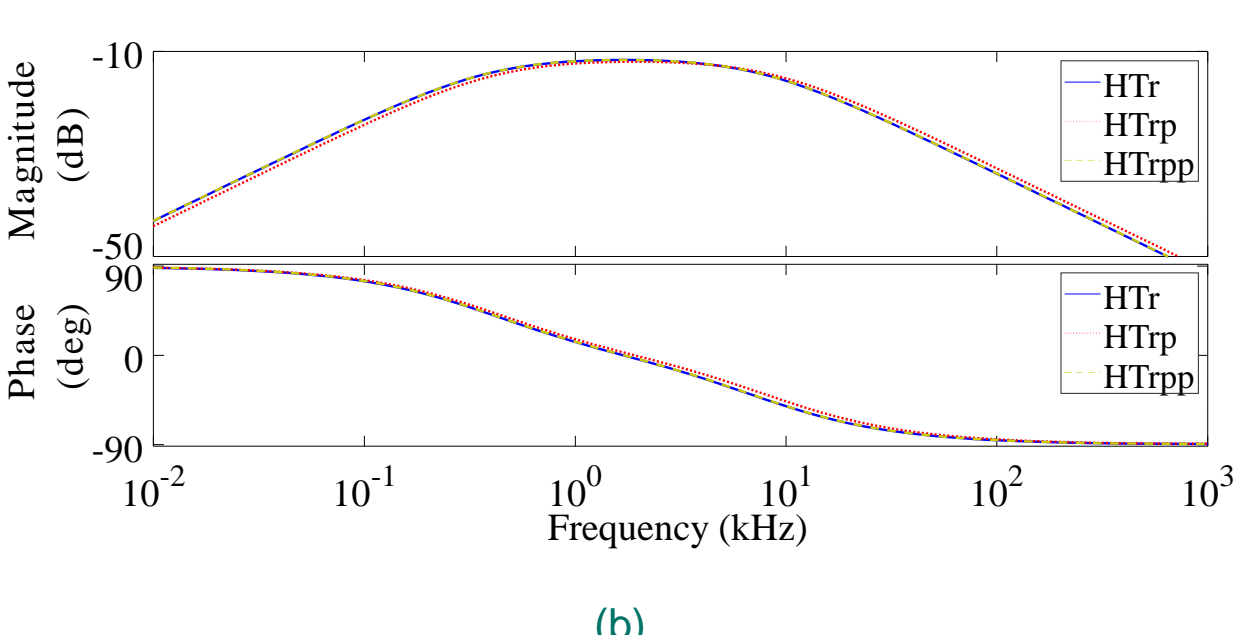

(b)

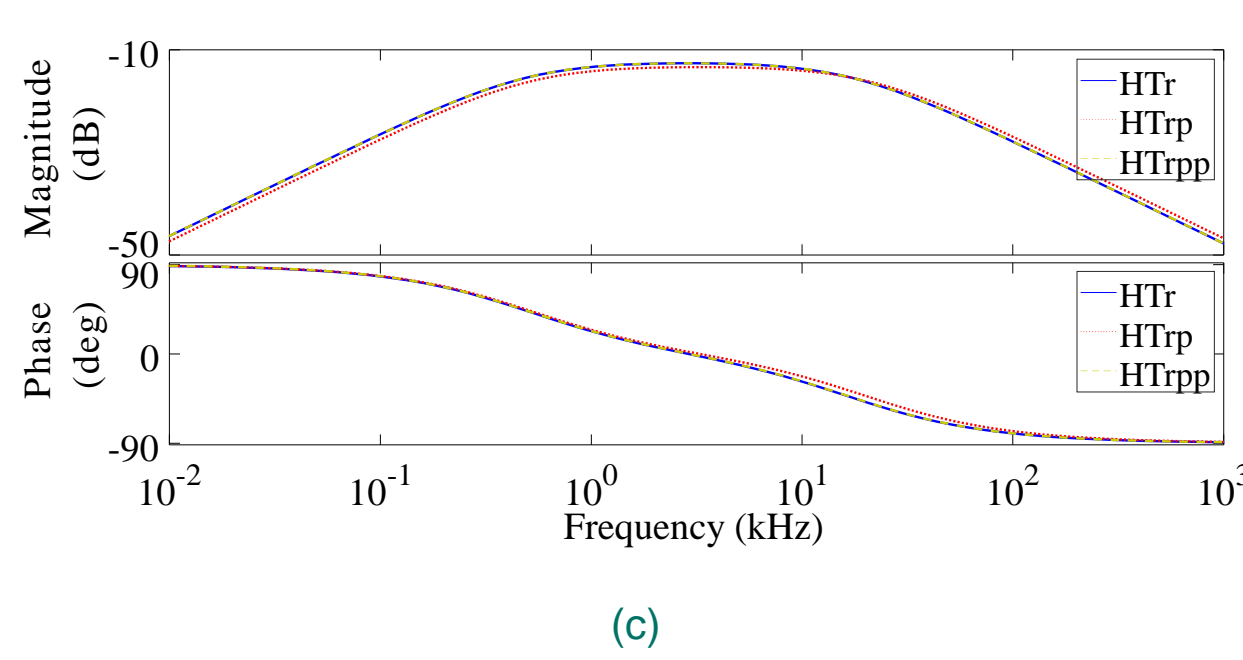

(c)

Fig. 12. Bode diagrams for different turn number ratios (a) 0.5, (b) 1, and (c) 2, where $H_{Tr}$ denotes the exact transformer ratio, $H_{Trp}$ denotes the turn number ratio, and $H_{Trpp}$ denotes the adjusted transformer ratio for transfer functions of transformer.

## IV. EXPERIMENTAL MEASUREMENT

To experimentally validate the proposed magnetic model and the modified transformer ratio formulation, the voltage transfer characteristics between the rotor and stator windings were measured. A prototype of the investigated rotary transformer was fabricated using the same geometrical dimensions as the small-size transformer listed in Table I, as shown in Fig. 13. The experimental setup is illustrated in Fig. 14. A digitally synthesized function generator with a frequency resolution of 0.1 Hz was used to supply the excitation signal, while the excitation voltage magnitude was regulated by an automatic gain control circuit. During the experiments, the excitation voltage and frequency were set to 2.5 V and 4 kHz, respectively. The tests were conducted under both no-load and loaded conditions to evaluate the influence of the transformer ratio adjustment. In the loaded case, a resolver with $R_L = 19\ \Omega$ and $L_L = 2.289$ mHwas used as the load.

The measured stator- and rotor-side voltages under different air-gap lengths and loading conditions are presented in Fig. 15. Furthermore, the corresponding results obtained from the 3D-FEA model and the equivalent circuit of Fig. 2, where the leakage and magnetizing inductances are calculated using the proposed magnetic model and the transformer ratio is determined according to (25), are shown in Figs. 16 and 17.

As summarized in Table IV, the proposed modeling approach shows good agreement with both experimental measurements and 3D-FEA results for air gaps of 0.6 mm and 1.2 mm. In the worst case, the prediction error of the proposed model is less than 7%. In contrast, when conventional modeling approaches reported in previous studies are employed, the deviation from the experimental results can reach approximately 42%.

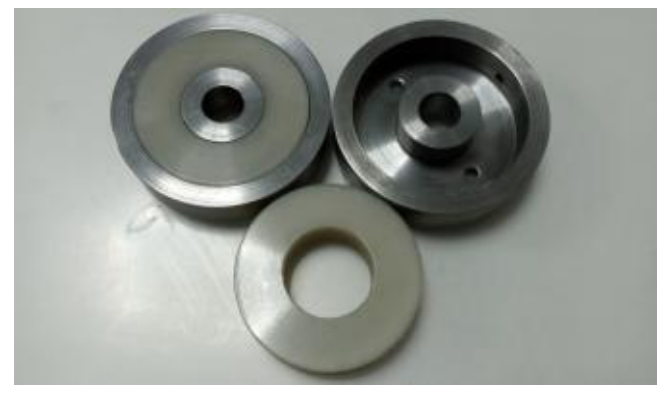

(a)

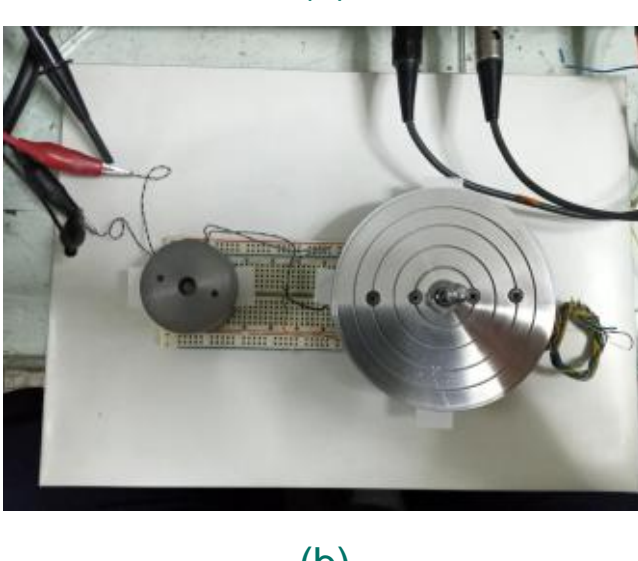

(b)

Fig. 13. The prototyped (a) rotary transformer has the same geometrical dimensions as the small-size transformer given in Table I, supplying a (b) resolver as a load.

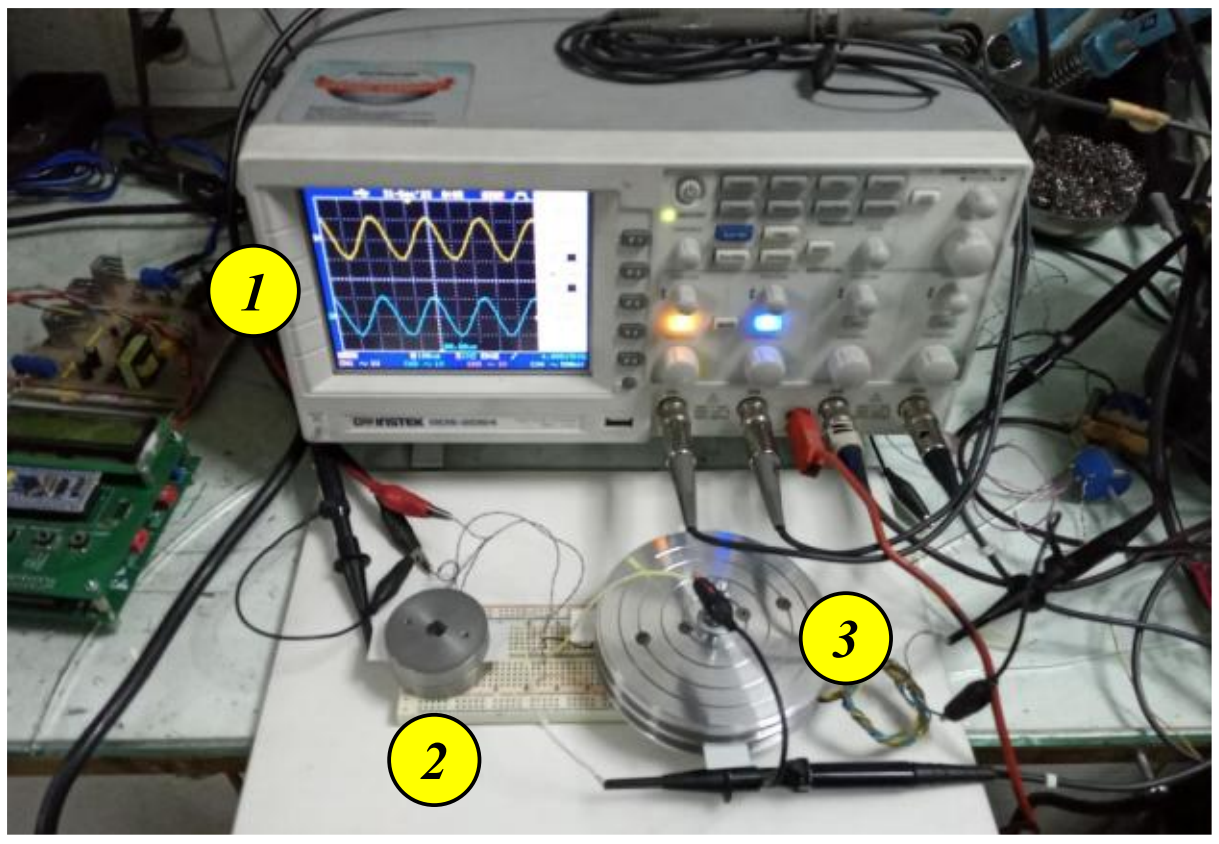

Fig. 14. Test bench with resolver load: (1) stator- and rotor-side voltages, (2) the prototyped rotary transformer, and (3) resolver as a load.

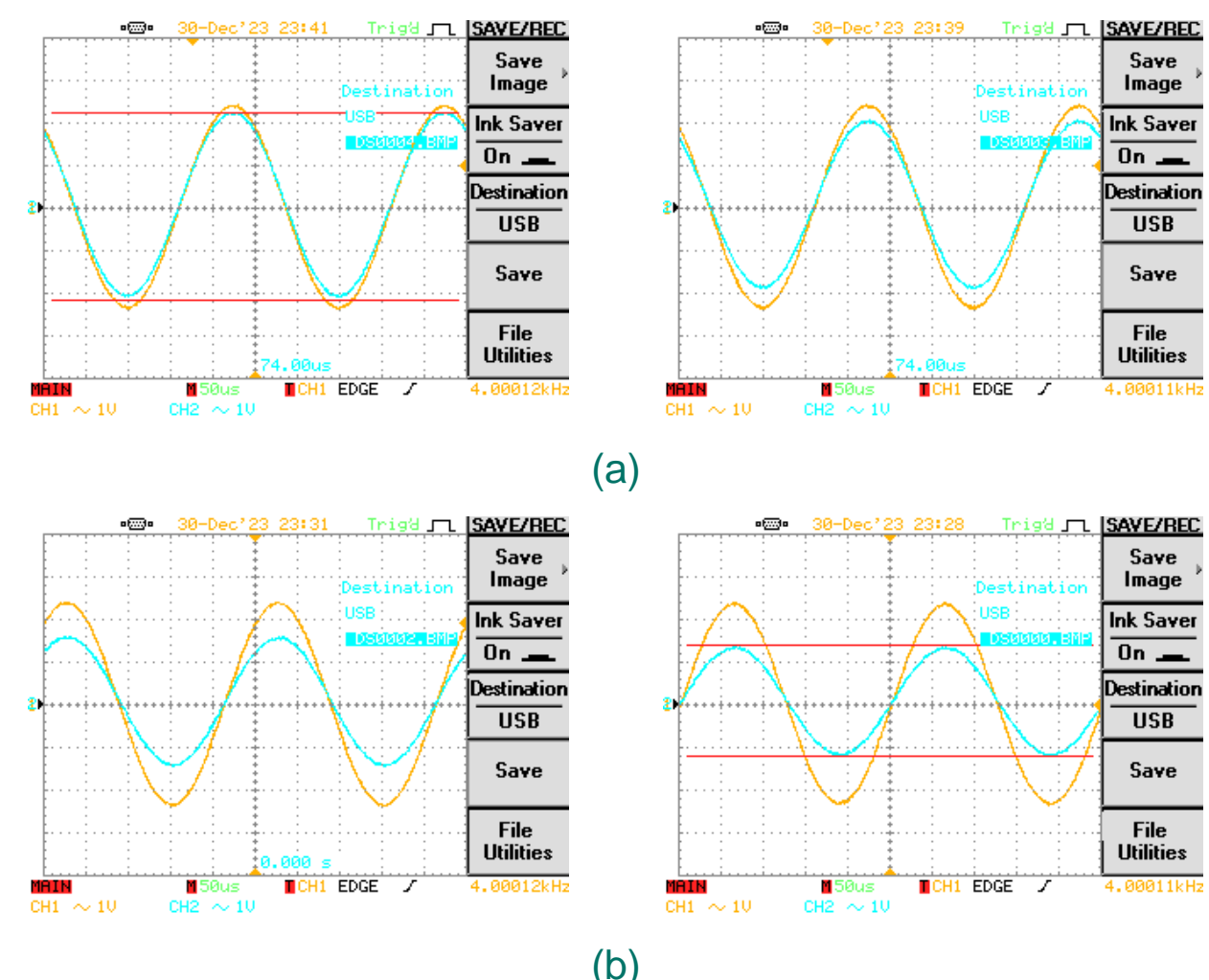

Fig. 15. The results of measurement, stator- and rotor-side voltages, for airgaps of 0.6mm (left) and 1.2mm (right): (a) No-load condition (b) resolver with $R_L = 19\ \Omega$ and $l_L = 2.289\ mH$ as a load.

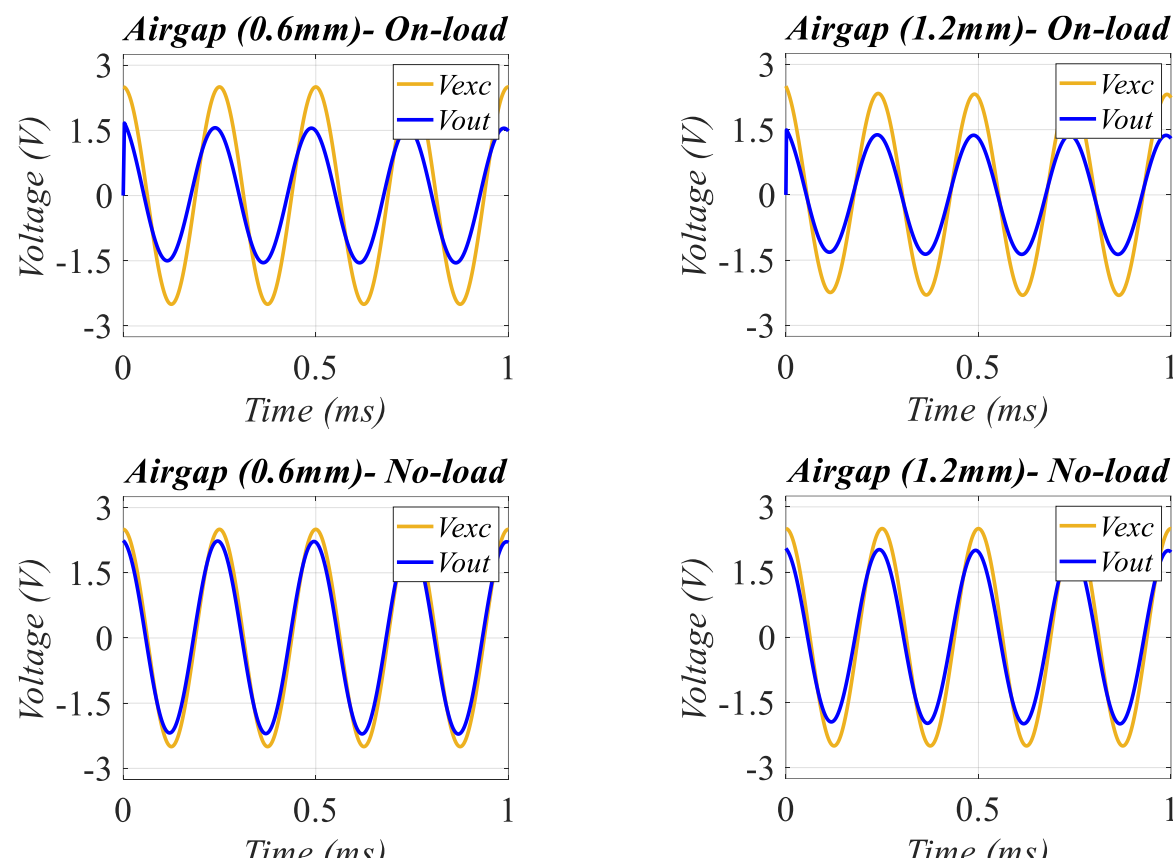

Fig. 16. The results of 3D-FEA, stator- and rotor-side voltages, for airgaps of 0.6mm and 1.2mm: on-load and no-load.

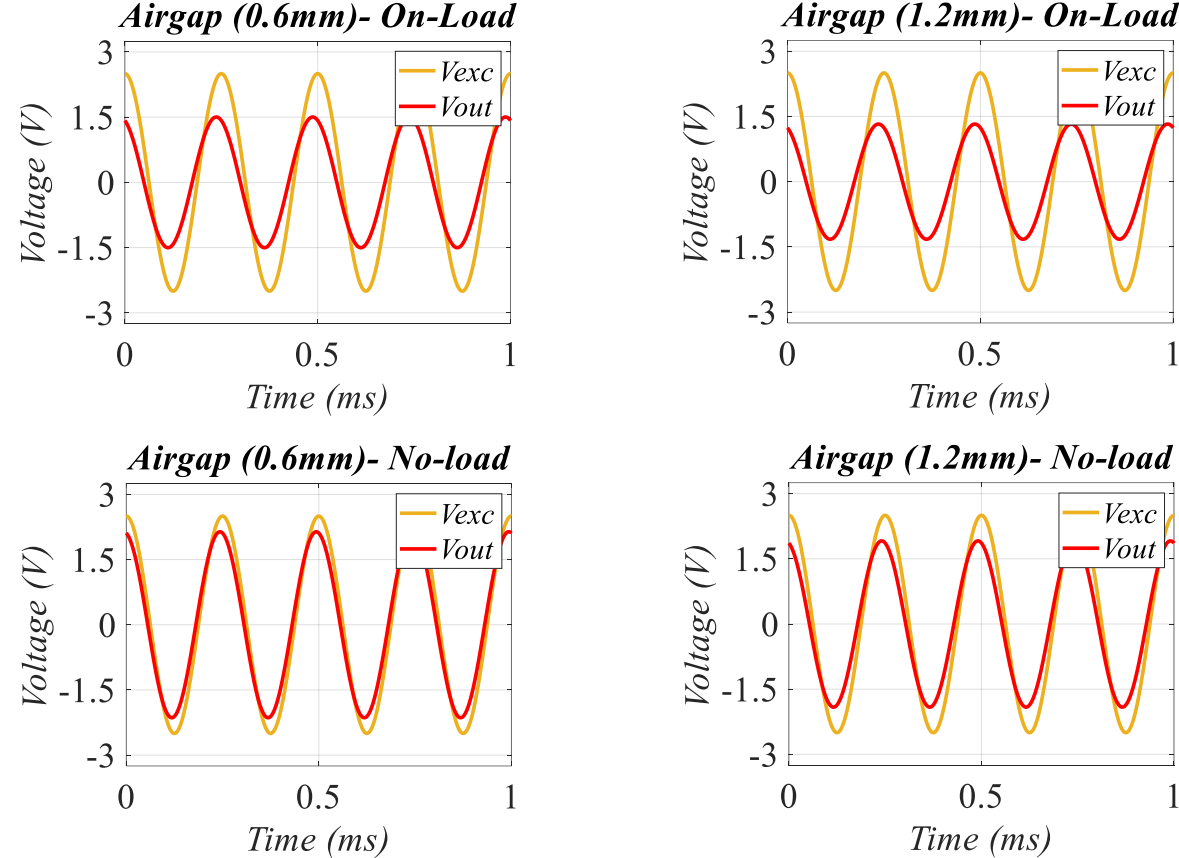

Fig. 17. The results of the circuit model in Fig. 2, including stator- and rotor-side voltages, based on the suggested magnetic model for the calculation of leakage and magnetization inductances, as well as the suggested transformer ratio adjustment, for air gaps of 0.6mm and 1.2mm: on-load and no-load.

## V. Conclusion

This paper presented an analytical modeling approach for miniature rotary transformers used in wound-rotor resolvers.

TABLE III
COMPARISON OF THE SUGGESTED METHODOLOGY, 3D FEA, AND EXPERIMENTAL MEASUREMENT

| Load | Airgaps | Rotor-Side Voltages | | | |
|---|---|---|---|---|---|
| | | The Standard Methodology | The Suggested Methodology | 3D FEA | Experimental Measurement |
| No-Load | 0.6 mm | 1.7 | 2.14 | 2.21 | 2.20 |
| | 1.2 mm | 1.17 | 1.91 | 1.99 | 2.00 |
| Resolver Load | 0.6 mm | 1.31 | 1.50 | 1.55 | 1.60 |
| | 1.2 mm | 0.94 | 1.32 | 1.37 | 1.40 |

Unlike conventional analyses that assume an ideal transformer behavior, the proposed model accounts for the non-ideal voltage transfer characteristics caused by significant leakage inductance in compact structures. A magnetic equivalent circuit was developed to derive the magnetizing and leakage inductances while considering flux fringing and air-gap effects. The obtained analytical expressions were validated through three-dimensional finite-element analysis and experimental measurements on a fabricated prototype under both no-load and resolver excitation conditions. The results demonstrated that the proposed model can accurately predict the secondary voltage of the rotary transformer, with a maximum error of about 7% compared with experimental results. In contrast, conventional modeling approaches may lead to errors of up to 42%. Therefore, the proposed formulation provides a more reliable tool for the analysis and design of miniature rotary transformers in compact resolver-based sensing systems.